\documentclass[prl,twocolumn]{revtex4}
\usepackage{bm}
\usepackage{graphicx}
\usepackage{amsmath}
\usepackage{eufrak}
\usepackage[utf8]{inputenc}
\usepackage[T1]{fontenc}
\usepackage{color}
\usepackage{upgreek}
\usepackage{subfigure}
\usepackage[unicode=true,colorlinks=true,citecolor=blue]{hyperref}
\usepackage{placeins}
\usepackage{lipsum}
\newcommand{\nix}[1]{}

\begin{document}

\title{High frequency impact ionization and nonlinearity of photocurrent induced by intense terahertz 
	radiation in HgTe-based quantum well structures}

\author{S. Hubmann$^1$, G.V. Budkin$^2$, A.P. Dmitriev$^2$, S. Gebert$^1$, V.V. Bel'kov$^2$, E.L. Ivchenko$^2$,  S. Baumann$^1$, M. Otteneder$^1$, D.A. Kozlov$^3$, 
	N.N. Mikhailov$^3$, S.A. Dvoretsky$^3$, Z.D. Kvon$^3$ and S.D. Ganichev$^1$}

\affiliation{$^1$Terahertz Center, University of Regensburg, 93040 Regensburg, Germany}

\affiliation{$^2$Ioffe Institute,
194021	St. Petersburg, Russia}

\affiliation{$^3$Rzhanov Institute of
Semiconductor Physics, 630090 Novosibirsk, Russia}

\begin{abstract}
We report on a strong nonlinear behavior of the photogalvanics and 
photoconductivity under excitation of HgTe quantum wells (QWs) by 
intense terahertz (THz) radiation. The increasing radiation intensity 
causes an inversion of the sign of the photocurrent and transition to its 
superlinear dependence on the intensity. The photoconductivity also 
shows a superlinear raise with the intensity. We show that the observed 
photoresponse nonlinearities are caused by the band-to-band \emph{light} 
impact ionization under conditions of a photon energy less than the 
forbidden gap. The signature of this kind of impact ionization is that 
the angular radiation frequency $\omega=2\pi f$ is much higher than the 
reciprocal momentum relaxation time. Thus, the impact ionization takes 
place solely because of collisions in the presence of a high-frequency 
electric field. The effect has been measured on narrow HgTe/CdTe QWs of 
5.7\,nm width; the nonlinearity is detected for linearly and circularly 
polarized THz radiation with different frequencies ranging from $f=0.6$ 
to 1.07\,THz and intensities up to hundreds of kW/cm$^2$. We demonstrate 
that the probability of the impact ionization is proportional to the 
exponential function, $\exp(-E_0^2/E^2)$, of the radiation electric 
field amplitude $E$ and the characteristic field parameter $E_0$. The 
effect is observable in a wide temperature range from 4.2 to 90\,K, with 
the characteristic field increasing with rising temperature.
\end{abstract}

\maketitle

\section{Introduction}
\label{introduction}

With the emergence of high-power pulsed terahertz sources such as  molecular lasers, free-electron lasers and few-cycle difference-frequency based terahertz systems there has been a surge in studies of intense terahertz excitation of semiconductors, see e.g.~\cite{Ganichev2005,Kampfrath2010,Kampfrath2013,Langer2016,Baierl2016,Banks2017,
Pietka2017,Hafez2018,Loon2018}.
High-intensity electromagnetic radiation of the terahertz range gives rise to a variety of novel nonlinear phenomena whose characteristic features are basically different from the corresponding effects at microwave frequencies as well as for the visible light.
In particular, high electric or magnetic fields of the THz radiation allow one a direct access to a number of low-energy elementary excitations such as phonons, plasma oscillations, spin waves, etc. or can drive the system into a non-perturbative regime of light-matter interaction. Moreover, the latest advances in terahertz technology made it possible to study nonlinear optical and opto-electronic phenomena on the femtosecond time scale with subcycle time resolution. Broadly speaking, experiments with powerful THz laser sources have potential to define limits of existing high frequency electronics, where the radiation field has a classical amplitude, and uncover new approaches in the development and application of future electronics at these frequencies. 

The intense  THz radiation can be used to convert topologically trivial HgTe QWs into a nontrivial 2D Floquet topological insulator as suggested theoretically by Lindner et al. \cite{Lindner2011}. This would result in an appearance of chiral edge states which can be proved via the generation of edge photocurrents induced by circularly polarized THz radiation. The latter has recently been demonstrated in Ref.~\cite{Dantscher2017} studying 2D topological insulators based on HgTe QW of 8\,nm width with inverted band ordering \cite{Bernevig2006,Koenig2007}.	
Being inspired by Ref.~\cite{Lindner2011} we have investigated photoresponses in 5.7\,nm thick QWs excited by monochromatic intense THz radiation under the conditions suggested in that work. By examining the intensity dependence of the photogalvanic current we have observed a strong nonlinearity resulting in a current sign inversion with increasing the radiation intensity $I$ from a fraction of W/cm$^2$ up to hundreds of kW/cm$^2$ and a superlinear behavior at high power.

Further investigation and analysis have demonstrated, however, that the observed effect is caused by the light impact ionization~\cite{Ganichev1986a}. This phenomenon is shown to cause the generation of electron-hole pairs by radiation with photon energy considerably smaller than the forbidden gap under the condition where the radiation angular frequency $\omega$ exceeds the reciprocal momentum relaxation time $\tau^{-1}$. Under this condition, the charge carriers acquire high energies solely because of collisions in the presence of a high-frequency electric field~\cite{Ganichev2005,Ganichev1986a,Keldysh1965}. Being primarily observed in bulk InSb crystals this effect was further demonstrated for very different three- and 
two-dimensional semiconductor systems~\cite{Ganichev1994,Markelz1996,Gaal2006,Wen2008,Hoffmann2009,Hafez2016,
Tarekegne2017,Oladyshkin2017}. A distinction of the light impact ionization reported in the present work is that it occurs in a system with the Fermi level larger than the forbidden gap. Thus, in contrast to all the previous works, the electron gas heating is needed in order to deplete the occupied states in the region of the conduction-band bottom rather than to increase the number of hot electrons with high energies exceeding the energy gap. As shown below, the experiments and theoretical analysis demonstrate that in our study the probability of impact ionization is proportional to $\exp(-E_0^2/E^2)$, where $E$ is the radiation electric field amplitude and $E_0$ is the characteristic field parameter. This dependence has been also confirmed by experiments on the THz radiation-induced photoconductivity showing that, in line with the theory, the characteristic field $E_0$ is proportional to the radiation frequency $\omega$. Lastly, we show that the observed nonlinearity in the photocurrent is caused by the interplay of the photogalvanic effect excited in the conduction and valence bands.

\section{Samples and methods}
\label{samples_methods}

\begin{figure}
	\centering
	\includegraphics[width=\linewidth]{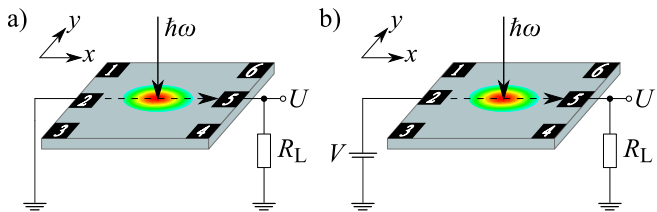}
	\caption{Panel (a): Setup scheme for a photocurrent measurement between contacts 2 and 5. The current is measured as a voltage drop $U$ across a load resistor. Panel (b): Setup scheme for a photoconductivity measurement between contacts 2 and 5. The signal is measured as a voltage drop $U$ along a load resistor $R_L$, while a bias voltage $V$ is applied.}
	\label{setup}
\end{figure}

The nonlinear phenomena described here were studied in HgTe/HgCdTe QWs grown on (013)-oriented GaAs substrates by molecular beam epitaxy \cite{Dvoretsky2010}. We used quantum wells 
with two barriers made of Hg$_{0.4}$Cd$_{0.6}$Te, each 30~nm thick. The width of the used QWs was 
$d=5.7$~nm yielding a non-inverted parabolic energy spectrum~\cite{Bernevig2006}. The structures have been grown on  4~$\upmu$m thick CdTe layers, which completely relax an initial strain caused by the lattice mismatch with the GaAs substrate. We used van-der-Pauw sample geometry with a size of 
$5\times 5$~mm$^2$. To measure the photoresponse, six ohmic contacts, four at the corners and two at the middle of opposite edges, have been fabricated. Additionally, Hall bar structures with a size of 
$6\times 50$~$\upmu$m have been prepared and used for magnetotransport measurements. From the latter measurements we obtained mobility $\mu$ and carrier density $n$ at $T= 4.2$~K being 
$\mu = 2.5 \times 10^4$~cm$^2$/(Vs) and $n= 3 \times 10^{11}$~cm$^{-2}$, respectively.

For optical excitation we used a high-power pulsed molecular gas THz laser~\cite{Ganichev2002,Olbrich2011,Olbrich2014} optically pumped by a tunable CO$_2$ laser~\cite{Ganichev2003}. Using CH$_3$F, D$_2$O, and NH$_3$ as active media, laser radiation with frequencies of 0.6, 0.77, and 1.07\,THz ($\hbar\omega=$ 2.5, 3.2, and 4.4~meV) were obtained. The laser operated in single pulse regime with a pulse duration of about 100\,ns and a repetition frequency of 1\,Hz. The radiation power $P$ has been measured by a
fast room temperature  photon drag detector made of $n$-type Ge crystals~\cite{Ganichev1985}. 
The laser beam had an almost Gaussian shape as measured by a pyroelectric camera~\cite{Ganichev1999}. The radiation was focused by a parabolic mirror to a spot diameter of about 
2.5~mm being smaller than the sample size.
This allowed us to distinctly illuminate the edges or the center of the sample. In the following, the latter case is referred to as ''bulk''.
The highest peak intensities obtained for these frequencies were 60, 80, and 200~kW/cm$^2$, respectively. 
Note that in the used laser the intensity varies from pulse to pulse by about 15\,\%.
The samples were placed in either an optical cold finger cryostat with TPX windows or an optical temperature-regulated continuous flow cryostat with quartz windows. While in the former case we were able to obtain the highest level of radiation intensity at highest frequencies used in the work, in the latter case we studied the temperature dependence of the nonlinear response at the lowest frequency. The measurement have been carried out in the temperature range from $T=$ 4.2 to 90\,K. In all measurements the samples were illuminated at normal incidence, see Fig.~\ref{setup}.

To vary the laser radiation intensity, we used either calibrated attenuators or crossed polarizers. In the latter case the linearly polarized laser radiation first passed through the wire grating polarizer. Rotation of this polarizer resulted in the decrease of the radiation intensity and the rotation of polarization state. The second polarizer, being at a fixed position, causes a further decrease of the radiation intensity and ensured that the radiation is always equally polarized. By this method, we obtained a controllable variation of the radiation intensity. To modify the radiation polarization state, crystal quartz quarter-wave plates were used. 
Rotating the plate by the angle $\varphi$ with respect to the laser polarization along the $x$-axis we changed the degree of circular polarization $P_\text{circ}$ according to 
\begin{equation} \label{circ}
P_{\rm circ} = \sin 2\varphi\:, 
\end{equation}
and two other Stokes parameters $P_L$ and $\tilde{P}_L$, defined as the degrees of linear polarization in the axes $x,y$ and the axes $x',y'$ rotated by 45$^{\circ}$, according to~\cite{Belkov2005}
\begin{equation} \label{lin}
P_L = \frac{\cos(4\varphi)+1}{2}\:,\: \tilde{P}_L = \frac{\sin(4\varphi)}{2}\:.
\end{equation}

Photocurrent and photoconductivity studied in this work have been measured using the setups shown in Figs.~\ref{setup}(a) and (b), respectively. The photocurrent was measured in \emph{unbiased} samples via a voltage drop $U$ across a load resistor $R_{\text{L}}$ amplified by a voltage amplifier and detected by a digital broad-band oscilloscope. The $dc$ photoconductivity was measured applying a bias voltage $V=0.3$\,V, see Fig.~\ref{setup}(b).  
This voltage was applied either as a $dc$ voltage or as a pulsed bias with a pulse length much longer than the laser pulse, so that the voltage bias is in a "quasi-$dc$"-regime with respect to the laser pulse. By subtracting the signals detected for positive and negative polarities of the bias the photoconductivity signal was extracted. From this signal the relative change of the conductivity $\Delta\sigma/\sigma$ has been calculated.

\section{Results}
\label{results}

\subsection{Photocurrent}
\label{Photocurrent_results}

\begin{figure}
	\centering
	\includegraphics[width=\linewidth]{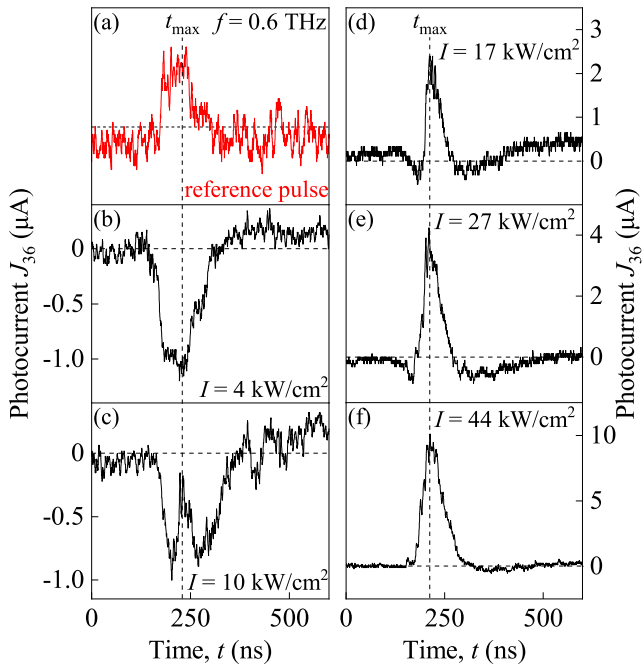}
	\caption{Panel (a): Reference laser pulse shape over time in arbitrary units. The signal is measured by the fast room temperature photon drag detector~\cite{Ganichev1985}. Panels (b-f): Time variation of the photocurrent response excited by the illumination of bulk with right-handed circularly polarized radiation at different intensities with a frequency of $f = 0.6$~THz and measured diagonally between contacts 3 and 6. At low intensities the current response follows the reference pulse with a negative amplitude, while at the higher intensities a nonlinear positive contribution is observed, which results in a change of sign of the photocurrent at approximately 10~kW/cm$^2$.}
	\label{pulse_pictures}
\end{figure}

We firstly describe the results of the photocurrent measurements.
Illuminating the edges or the bulk of the HgTe QWs with linearly or circularly polarized radiation we observed complex dynamics of the photoresponse at high radiation intensities. Figure~\ref{pulse_pictures} demonstrates an example of the genesis of the photoresponse with rising intensity detected for right-handed circularly polarized radiation with $f=0.6$~THz. At relatively low incident intensities the signal pulse is negative for any moment in time and its temporal shape repeats that of the reference laser pulse, Figs.~\ref{pulse_pictures}(b) and (a), respectively. As the intensity is increased, the signal dynamic changes. Now, at first the negative signal amplitude increases, but then it rapidly drops to zero at the maximum of the excitation pulse ($t = t_{\rm max}$), rises again and finally vanishes following the excitation pulse, see Fig.~\ref{pulse_pictures}(c). On further increasing the intensity, the photosignal changes its sign at some value $I_{\rm inv}$ and becomes positive at the maximum of the excitation pulse. At even higher intensities the positive part of the signal pulse dominates the photoresponse, Figs.~\ref{pulse_pictures}(d)-(f). The complicated temporal structure of the signal can be understood assuming that it emerges simply due to the change of sign of the photocurrent as a function of radiation intensity. On this assumption, at the rising edge of the laser pulse the intensity increase upon time causes the dynamic inversion of the photoresponse, whereas at the falling pulse edge the intensity decreases and the signal dynamic mirrors. Consequently, the peak of the photocurrent occurs at the time $t_{\rm max}$ of the laser intensity maximum.

\begin{figure}
	\centering
	\includegraphics[width=\linewidth]{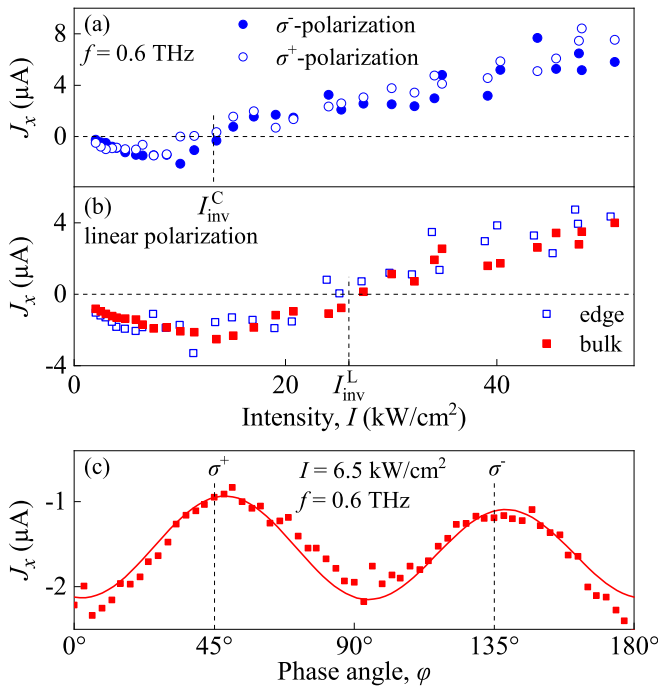}
	\caption{Panel (a): The intensity dependences of the edge photocurrent excited by right-handed (open circles) and left-handed (full circles) circularly polarized radiation with a frequency of $f = 0.6$~THz. The signal was picked up from  contacts 1 and 6 under illumination of the corresponding sample edge. The change of sign and the nonlinear raise of the photocurrent is clearly seen, while there is no difference within the measuring accuracy between signals for the left- and right-handed excitation polarizations. 
		Panel (b): Dependences of the photocurrent at the edge (blue, picked up from contacts 1 and 6) and in the bulk (red, picked up from contacts 2 and 5) excited by linearly polarized radiation with a frequency of $f = 0.6$~THz  on the intensity $I$. Panel (c): Dependence of the photocurrent induced in the bulk on the quarter-wave plate angle $\varphi$ measured at the intensity of 	$I= 6.5$~kW/cm$^2$ for the radiation frequency of $f=0.6$~THz. The solid line shows a fit after Eq.~\eqref{helicity_eqn} with the fitting parameters $J_0= - 1.0\,\upmu$A, $J_1=-1.1\,\upmu$A, $J_2=-0.3\,\upmu$A, and $J_3=0.1\,\upmu$A.
	}
	\label{bulkedge}
\end{figure}

\begin{figure}
	\centering
	\includegraphics[width=\linewidth]{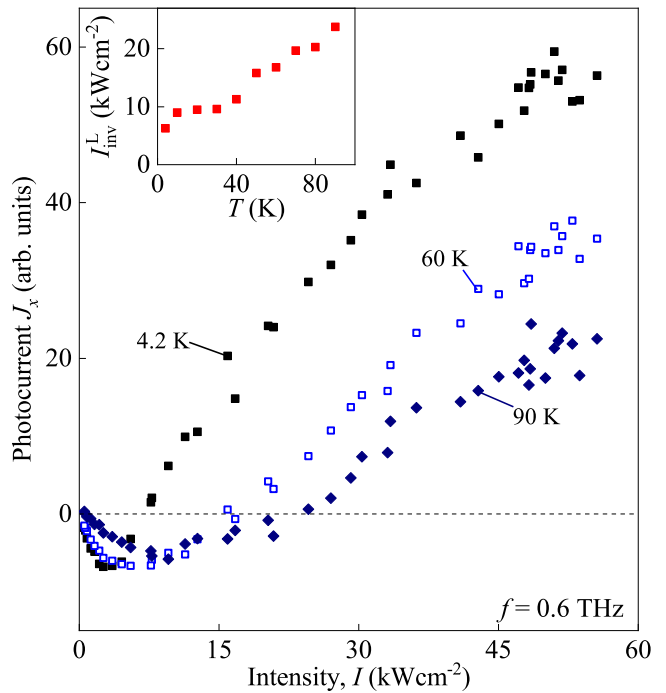}
	\caption{Intensity dependences of the photocurrent excited by linearly polarized radiation with a frequency of $f = 0.6$~THz for the temperatures of $T=4.2$, 60, and 90\,K. The signal was picked up from contacts 2 and 5 under illumination of the sample bulk. The inset shows the temperature dependence of the inversion intensity $I_{\text{inv}}^{\text{L}}$, which shifts to higher intensities at higher temperatures.}
	\label{temperature}
\end{figure}

Figure~\ref{bulkedge} shows the intensity dependence of the photocurrent measured at time $t = t_{\rm max}$. The open and full circles in Fig.~\ref{bulkedge}(a) are obtained for illumination of the sample edge by circularly polarized radiation with opposite helicities. The data reveal almost no difference in the behavior of the photocurrents excited by the $\sigma^+$ and $\sigma^-$-polarizations: The both dependences show the inversion of sign at the intensity of about $I_{\rm inv}^{\text C}=15$~kW/cm$^2$.
The independence of the radiation helicity is additionally confirmed by studying the current variation with the phase angle $\varphi$ (not shown). Similar intensity dependence, but with the higher value of $I_{\rm inv}^{\text L}\approx 25$~kW/cm$^2$, has been observed for excitation with linearly polarized radiation, empty squares in Fig.~\ref{bulkedge}(b). Studying the photoresponse in a wide temperature range from 4 to 90\,K we observed that qualitatively the nonlinear behavior remains the same, see Fig.~\ref{temperature}. The only difference is that the intensity when the sign inversion occurs increases with increasing temperature.

The insensitivity of the photoresponse to the radiation helicity demonstrates that the nonlinear photocurrent is not caused by helical edge photocurrents as a consequence of a transition from topologically trivial to nontrivial phase. Moreover, the current is not even generated at the sample edges: Shifting the beam spot into the sample bulk (illumination of the sample center) and measuring the photocurrent across the sample we also observed the photocurrent sign change with rising the intensity.
Figure~\ref{bulkedge}(b) shows the comparison of the photocurrent excited by linearly polarized radiation at the edge (blue empty squares) and in the bulk (red full squares).
The data reveal no measurable difference between the bulk and edge results excluding edge mechanisms as a driving force for the observed nonlinear photocurrent. In the following we will present only the data obtained illuminating the bulk of the sample. Furthermore, we have observed that switching of the radiation helicity from the right- to left-handed polarized radiation does not substantially affect the nonlinearity (not shown).

The photocurrent at low intensity is most likely caused by photogalvanic effects, well known for HgTe QWs \cite{Wittmann2010,Dantscher2017}. This is supported by the polarization dependence detected at low intensities, see  Fig.~\ref{bulkedge}(c), which is described well by
\begin{eqnarray} \label{N1}
\label{helicity_eqn}
J_x &=& J_0 + J_1 (|e_x|^2 - |e_y|^2) + J_2\ {\rm Re}(2 e^*_x e_y) + J_3 P_{\rm circ}\nonumber \\ &=& J_0+J_1 P_L(\varphi) +J_2 \tilde{P}_L(\varphi)  +J_3 P_{\rm circ}(\varphi)\:,
\end{eqnarray}
where ${\bm e}$ is the polarization unit vector and the Stokes parameters are defined by Eqs.~(\ref{circ}) and (\ref{lin}). The fitting parameters $J_0$, $J_1$, and $J_2$ correspond to different contributions of the linear photogalvanic effect (LPGE) and the parameter $J_3$ describes the circular photogalvanic effect.  Figure~\ref{bulkedge}(c) reveals that the photocurrent is dominated by the LPGE.



As we show below (Sec.~\ref{discussion}) the observed strong nonlinearity resulting in the change of the direction of the photogalvanic current with rising intensity is caused by the generation of electron-hole pairs by the THz radiation as a result of light impact ionization. To study the rate of electron-hole pair generation we have measured the THz radiation induced change of conductivity $\Delta\sigma$.

\subsection{Photoconductivity}
\label{Photoconductivity_results}

\begin{figure}
\centering
\includegraphics[width=\linewidth]{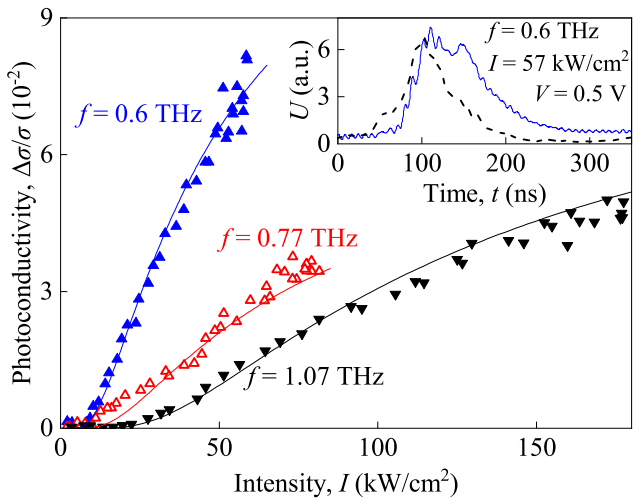}
\caption{Dependences of the normalized photoconductivity $\Delta\sigma/\sigma$ excited by linearly polarized radiation in the bulk of the QW for the radiation frequencies of 0.6 (blue triangles), 0.77 (red triangles), and 1.07~THz (black triangles) on the intensity $I$. The signal is measured for contacts 2 and 5 using the setup shown in Fig.~\ref{setup}(b). Solid lines present the corresponding fits after Eq.~\eqref{exp_eqn} [see also theoretical Eq.~\eqref{exp_theory}] with the fitting parameters $A$ and $I_0$. A nonlinear raise of the signal is clearly seen for all frequencies.
Inset shows the kinetics for an exemplary photoconductivity pulse (solid line) measured in the bulk of the samples for linearly polarized radiation with a frequency of 0.6~THz and an intensity of 57~kW/cm$^2$. The dashed line shows the reference laser pulse.}
\label{PC_Intensity}
\end{figure}

\begin{figure}
	\centering
	\includegraphics[width=\linewidth]{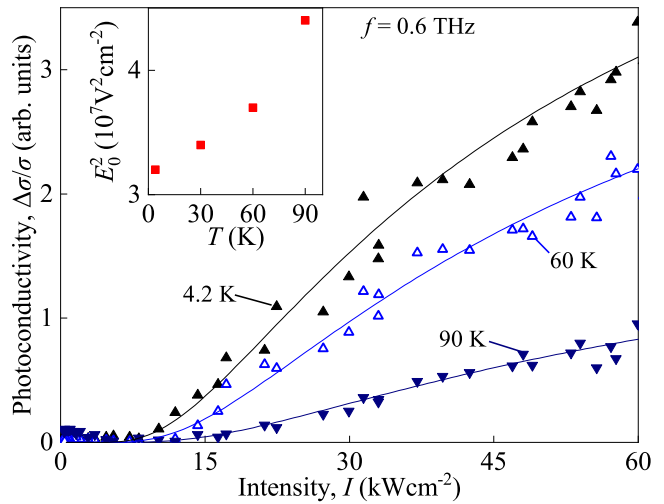}
	\caption{Intensity dependences of the normalized photoconductivity $\Delta\sigma/\sigma$ excited by linearly polarized radiation with $f=0.6$\,THz in the bulk of the QW. The signal was measured for three temperatures of $T=4.2$, 60, and 90\,K for the contacts 2 and 5 using the setup shown in Fig.~\ref{setup}(b). Solid lines show corresponding fits after Eq.~\eqref{exp_eqn} [see also theoretical Eq.~\eqref{exp_theory}] with the fitting parameters $A$ and $I_0$. A nonlinear raise of the signal is clearly seen for all frequencies. Inset shows the temperature dependence of the fitting parameter $E_0^2\propto I_0$.}
	\label{temp2}
\end{figure}

A positive photoconductivity response
was detected for all measured frequencies. In contrast to the photocurrent, the photoconductivity response time is twice longer than the laser pulse, see inset in Fig.~\ref{PC_Intensity}. The observed increase of the sample's conductivity together with the photoresponse kinetics reveals that it is caused by the generation of electron-hole pairs,
in spite of the fact that the photon energy is smaller than the band gap, $\hbar\omega < \varepsilon_\text{g}$. 
Note that the detected 
response times correspond to the recombination of electron-hole pairs known for HgTe QWs~\cite{Morozov2012}.
The characteristic intensity dependencies of the photoconductivity are shown in Fig.~\ref{PC_Intensity} for illumination of the bulk by linearly polarized radiation of different frequencies. 
Figure~\ref{temp2} shows the photoconductivity as a function of the radiation intensity for the frequency of 0.6\,THz and three temperatures in the range between 4 and 90\,K.

\begin{figure}
	\centering
	\includegraphics[width=\linewidth]{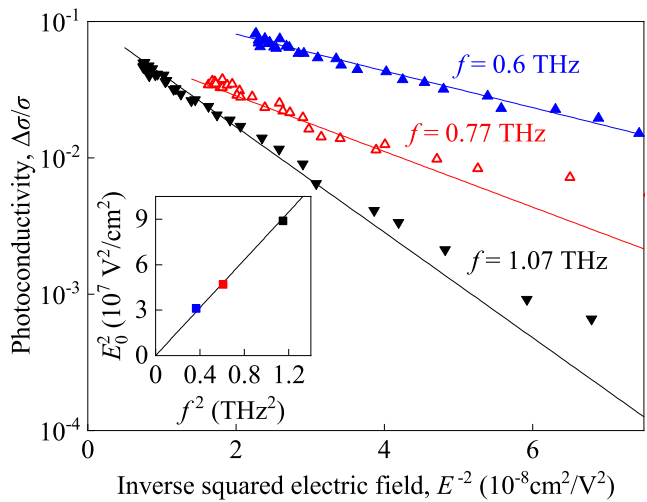}
	\caption{Logarithmic plot of the relative photoconductivity $\Delta\sigma/\sigma$ excited by linearly polarized radiation in the bulk of the sample on the inverse squared electric field $E^{-2}$. Data are shown for the radiation frequencies of 0.6\,(blue triangles), 0.77\,(red triangles), and 1.07~THz\,(black triangles) and have been measured between the contacts 2 and 5. Solid lines present the corresponding fits after Eq.~\eqref{exp_eqn} [see also theoretical Eq.~\eqref{exp_theory} and \eqref{eq:6}.] 
	 with fitting parameters $A$ and $E_0^2$. Inset shows a dependence of the fitting parameter $E_0^2$ extracted from the corresponding fits from panel (a) on the squared radiation frequency $f^2$.}
	\label{PC_Logarithmic}
\end{figure}

The results clearly show a superlinear raise of the photoconductive signal and reveal that the nonlinearity decreases with increasing the radiation frequency. The data can be well described by
\begin{equation}
\label{exp_eqn}
\frac{\Delta\sigma}{\sigma}=A\cdot\exp\left(-\frac{E_0^2}{E^2}\right)=A\cdot\exp\left(-\frac{I_0}{I}\right)
\end{equation}
with the prefactor $A$ and the characteristic electric field $E_0$ as fitting parameters.
Replotting the data in a half-logarithmic plot as a function of the inverse squared electric field $E^{-2}\propto I^{-1}$ we see that the above equation describes  the data well, see Fig.~\ref{PC_Logarithmic}(a). Extracting $E_0$ from the slope of the fit lines we obtained that it changes according to $E_0\propto f$, see inset in Fig.~\ref{PC_Logarithmic}(b). We note that at low intensities and high frequencies a deviation from Eq.~\eqref{exp_eqn} occurs. This is attributed to a contribution of $\mu$-photoconductivity (bolometric photoresponse)~\cite{Ganichev2005}. 

To summarize the experimental part, we have shown that the intense terahertz excitation of HgTe QW structures results in a strongly nonlinear photoresponse which is caused by the generation of electron-hole pairs despite of the fact that the photon energy is substantially smaller than the energy gap. Electron-hole pair generation results (i) in a nonlinear photoconductivity, which scales as $\exp\left(-E_0^2/E^2\right)$ with $E_0\propto f$, and (ii) in a sign-inversion of the photogalvanic currents at high intensities.
The fact that for smaller photon energies the nonlinearity is observed at substantially smaller radiation intensities, see Fig.~\ref{PC_Logarithmic}, together with the lack of the linear-circular dichroism (not shown), excludes multiphoton processes as an origin of the nonlinearities. The observed electric field and frequency dependencies of the photoconductivity give a hint that the observed nonlinearity is caused by the light impact ionization, previously observed in bulk InSb and InAs-based materials~\cite{Ganichev1984,Ganichev1986a,Ganichev1994,Markelz1996}.


\section{Discussion}
\label{discussion}

\begin{figure}[htbp]
	\centering
	\includegraphics{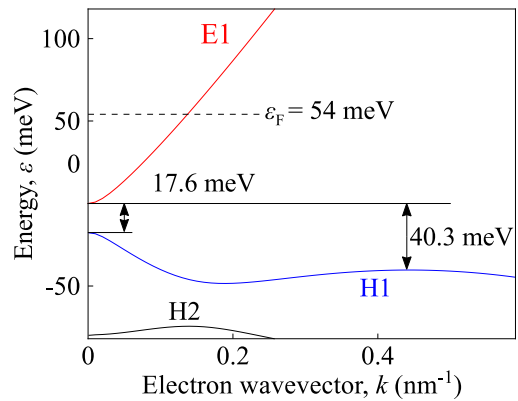}
	\caption{Calculated energy dispersion for a quantum well with a thickness of 5.7~nm using the $\bm k \cdot \bm p$ Hamiltonian from Ref.~\cite{Novik2005}. Here, the band gap between the electron subband E1 (red) and the hole  subband H1 (blue) is $\varepsilon_\text{g}= 17.6$~meV. The dashed line indicates the Fermi level.
		}
	\label{57spectrum}
\end{figure}

Now we discuss the microscopic origin of the observed nonlinear photoconductivity and photocurrent. 
First of all we discuss the band structure of the studied samples which is required to understand the mechanism of the electron-hole pair generation. 
We calculate electron states and dispersion of the (013)-oriented 5.7~nm thick HgTe QW in the 8-band model including the conduction ($\Gamma_6$), valence ($\Gamma_8$), and spin-orbit split-off bands ($\Gamma_7$). The effective $\bm k$$\cdot$$\bm p$ 
Hamiltonian in the Kane model is taken from Ref.~\cite{Novik2005}. 
Figure~\ref{57spectrum} shows the energy dispersion for three subbands $E1$, $H1$, and $H2$. 
Here, at  $k_{\|}=0$ the $|E1,\pm 1/2\rangle$ quantum well subband state
is formed from the linear combination of the $|\Gamma_6, \pm 1/2\rangle$ and $|\Gamma_8, \pm 1/2\rangle$ states while $|H1\rangle$ and $|H2\rangle$ states are the first and second levels of the size quantization of heavy holes, respectively. 
Figure~\ref{57spectrum} reveals that the band gap in the studied QWs is $\varepsilon_\text{g}= 17.6$~meV. From magnetotransport measurements we obtained that the Fermi level is $\varepsilon_\text{F}= 54$~meV, i.e., is substantially larger than the band gap.

\subsection{Photoconductivity}

In all previous works aimed to impact ionization, the typical values of the electron energy $\varepsilon$ in the equilibrium state are much lower than the threshold energy of impact ionization $\varepsilon_\text{i}$. 
Therefore, to achieve an appreciable rate of generation of the pairs, a sufficiently strong heating of the electron gas is necessary, leading to an increase in the number of electrons with high energy. 

In the samples studied here at rather high electron concentrations, the Fermi energy $\varepsilon_\text{F}$ exceeds $\varepsilon_\text{i}$ already in the equilibrium. In our case $\varepsilon_\text{i}$ is about half as much as $\varepsilon_\text{F}$. Thus the generation of pairs is limited not by the lack of electrons with high energy, but rather by the small number of unoccupied levels in the low-energy region. 
In such a situation, heating is needed
in order to deplete the occupied levels in the region of relatively low energies. 

We consider the dominant mechanisms of electron scattering in the discussed polar HgTe crystals at low temperatures tp be spontaneous emission of polar optical phonons with the energy $\varepsilon_0 = \hbar\omega_0$ and elastic scattering by point impurities with time $\tau_i(\varepsilon)$. We also take into account that for the experiments discussed above the condition $\omega\tau>1$ is fulfilled. Indeed, from  the width of the cyclotron resonance
we obtained 
$\omega\tau\approx 2$ for the studied range of frequencies.

For a radiation electric field polarized along the $x$-axis relevant to the experiment, the kinetic equation has the form 
%
\begin{equation}
\begin{split}
\frac{\partial f\left(\bm{p},t\right)}{\partial t}+eE\,\cos{(\omega t)}\,\frac{\partial f\left(\bm{p},t\right)}{\partial {p}_{x}}=\\-\frac{f\left(\bm{p},t\right)-{f}_{0}\left(\varepsilon ,t\right)}{{\tau }_{i}\left(\varepsilon \right)}-S{t}_{-}^{\rm ph}{\left\{f\right\}+S{t}_{+}^{\rm ph}}{\left\{f\right\}}.
\label{eq:1}
\end{split}
\end{equation}
Here, $f\left(\bm{p},t\right)$ is the distribution function depending on electron momentum $\bm{p}$ and time $t$, $e$ is the electron charge, $p(\varepsilon)$ and $v(\varepsilon)=\partial\varepsilon/\partial p$ are the momentum and velocity of an electron with the energy $\varepsilon$; 
$f_0(\varepsilon,t)$ is the part of the distribution function that is independent of  the direction of the electron momentum $\bm{p}$; 
$\tau_i (\varepsilon) = \tau_{i0}g(0)/g(\varepsilon)$, where $\tau_{i0}$  
is scattering time at the bottom of the conduction band
\cite{Gantmakher1988,footnotescattering};
 $g(\varepsilon) = p(\varepsilon) / v(\varepsilon)\pi\hbar^2$ is the density of states in the conduction band.
The combination $St_{+}^{\rm ph} \left\{f\right\}-St_{-}^{\rm ph} \left\{f\right\}$ describes the collision integral due to interaction with optical phonons. Separately,  $St_{+}^{\rm ph} \left\{f\right\}$ ($St_{-}^{\rm ph} \left\{f\right\}$) describes the number of electrons entering (leaving) the state with momentum $\bm{p}$ in an unit of time due to emission of phonons.

For the considered dominant mechanisms of electron scattering, the distribution function $f(\bm{p},t)$ is almost isotropic, i.e. $f(\bm{p},t)=f_0(\varepsilon)+f_1 (\bm{p},t)$, where $f_1 (\bm{p},t)$ is a small anisotropic correction. We assume that the electron gas heating is sufficiently strong, so that the effective electron temperature exceeds the phonon energy and, therefore, the function $f_0 (\varepsilon)$ only slightly changes on the energy scale of the order of $\varepsilon_0$. By that for $\varepsilon > \varepsilon_0$, $f_0(\varepsilon)$ is given by the balance equation 
%
\begin{equation}
D\left(\varepsilon \right)\frac{\partial {f}_{0}}{\partial \varepsilon }+\frac{{\varepsilon }_{0}}{{\tau}_{\rm ph}{\left(\varepsilon \right)}}{f}_{0}\left(\varepsilon \right)\left[1-f_{0}\left(\varepsilon \right)\right]=0
\label{eq:2}
\end{equation}
\begin{equation*}
D\left(\varepsilon \right)=\frac{{e}^{2}{E}^{2}{v}^{2}\left(\varepsilon \right)}{4{\omega }^{2}{\tau}_{i}\left(\varepsilon \right)},\:\frac{1}{{\tau}_{\rm ph}\left(\varepsilon \right)}=\frac{4\pi {\varepsilon }_{0}{e}^{2}g\left(\varepsilon \right)}{{\bar{\epsilon}} {p}\left(\varepsilon \right)},
\end{equation*}
where $1/\bar{\epsilon}=1/\epsilon_\infty - 1/\epsilon_0$; 
$\epsilon_\infty$ and $\epsilon_0$ are high- and low-frequency dielectric permittivities;
$D\left(\varepsilon \right)$ is the diffusion coefficient of electrons in the energy space, and $\tau_{\rm ph}(\varepsilon)$ is the characteristic time of phonon emission. 
The first term in Eq.~\eqref{eq:2} describes the heating of the electron gas by the electric field of the electromagnetic wave, whereas the second term corresponds to the energy losses due to emission of phonons.

The solution of Eq.~\eqref{eq:2} has the form

\begin{equation}
{f}_{0}\left(\varepsilon \right)=\frac{1}{1+\mathrm{exp}\left[-L\left(\varepsilon \right)\right]},\quad L\left(\varepsilon \right)=\underset{\varepsilon }{\overset{{\varepsilon}_{\text{E}}}{\int }}\frac{{\varepsilon }_{0}}{D\left({\varepsilon }^{\prime }\right){\tau}_{{\rm ph}}\left({\varepsilon }^{\prime }\right)}d{\varepsilon }^{\prime },
\label{eq:3}
\end{equation}

%
where $\varepsilon_\text{E}$ is determined from \mbox{$n=\underset{0}{\overset{{\infty}}{\int }} f_0(\varepsilon)g(\varepsilon)d\varepsilon$}, i.e. normalization by the density $n$.
 
For the considered condition of a slowly changing function $f_0 (\varepsilon)$ on the energy scale of the order of $\varepsilon_0$ the value $1-f_0(\varepsilon)$ also varies only slightly with energy $\varepsilon$. 
At $f_0(\varepsilon)$ close to unity,
$1-f_0(\varepsilon)$ is proportional to $\mathrm{exp}[-L(\varepsilon)]$. The above condition is fulfilled for $\mid L(\varepsilon+\varepsilon_0)-L(\varepsilon)\mid\:\ll 1$, which as it follows from Eq.~\eqref{eq:3} is equivalent to $\varepsilon_0^2\ll D\tau_{\rm ph}$.
For the opposite inequality, $\varepsilon<\varepsilon_0$, the distribution function obeys the equation
%
%
\begin{equation}
	\label{eq:4}
	\begin{split}
	-\frac{1}{g\left(\varepsilon \right)}\frac{\partial }{\partial \varepsilon}\left[g\left(\varepsilon \right)D\left(\varepsilon \right)\frac{\partial {f}_{0}}{\partial \varepsilon }\right]=\\\frac{1}{{t}_{\rm ph}^{+}\left(\varepsilon \right)}{f}_{0}\left(\varepsilon +{\varepsilon }_{0}\right)\left[1-{f}_{0}\left(\varepsilon \right)\right],
	\end{split}
\end{equation}
	where
 $f_0(\varepsilon+\varepsilon_0)$ is given by Eq.~\eqref{eq:2} and ${t}_{\rm ph}^{+}$ is given by Eq.~(\ref{X}) in the Appendix.

While it is not possible to solve this equation analytically, under the conditions $\varepsilon_0^2\ll D\tau_{\rm ph}$ and \mbox{$f_0(\varepsilon)\approx1$} relevant to the considerate case the right part of Eq.~\eqref{eq:4} becomes vanishingly small and, therefore, the distribution function in the region $\varepsilon<\varepsilon_0$ can be considered as almost constant. It can be calculated from Eq.~\eqref{eq:3} taking into account $\varepsilon=\varepsilon_0$. For a not so strong electron gas heating, so that the effective electron temperature is smaller than the Fermi energy, the distribution function is close to unity in the whole range of energies $0<\varepsilon<\varepsilon_\text{E}$. By contrast, at high energies $\varepsilon\gg\varepsilon_\text{E}$, it approaches zero. Consequently one obtains that $n\approx\underset{0}{\overset{{\varepsilon_\text{E}}}{\int }}g(\varepsilon)d\varepsilon$, which for $\varepsilon_\text{E}$ yields the value equal to the Fermi energy in equilibrium. This is because for the degenerated electron gas the latter is determined by the same integral.

The ionization rate is determined, as stated before, by the number of unoccupied levels in the low-energy region, i.e. by $1-f_0(\varepsilon)\approx\mathrm{exp}[-L(\varepsilon)]$. The exact value of the lower limit of integration in the expression for $L(\varepsilon)$ depends on the characteristics of the elementary act of impact ionization unknown to us, and therefore we shall consider it to be an adjustable parameter $\varepsilon^\ast$.
Subsequently, field and frequency dependences of the number of generated pairs are determined by the exponent $\exp[-L^\ast]$, where
\begin{equation}
{L}^{\ast}=\underset{{\varepsilon}^{\ast}}{\overset{{\varepsilon }_\text{F}}{\int }}\frac{{\varepsilon }_{0}}{D\left(\varepsilon \right){t}_{\rm ph}\left(\varepsilon \right)}d\varepsilon.
\label{eq:5}
\end{equation}

Within the Kane model, the energy is described by $\varepsilon(p)=\sqrt{\varepsilon_\text{g}^2 / 4+p^2\varepsilon_\text{g} / 2m}-\varepsilon_\text{g}/2$, which yields $g(\varepsilon)=m(2\varepsilon+\varepsilon_\text{g}) / \pi\hbar^2\varepsilon_\text{g}$, $p(\varepsilon)=\sqrt{2m\varepsilon(\varepsilon+\varepsilon_\text{g})\varepsilon_\text{g}^{-1}}$ and $v^2(\varepsilon)=2\varepsilon(\varepsilon+\varepsilon_\text{g})\varepsilon_\text{g} / m(2\varepsilon-\varepsilon_\text{g})^2$.
 Here $m$ is the effective mass of the electron at the bottom of the conduction band. Substituting these expressions into Eq.~\eqref{eq:2} 
 we obtain the rate of ionization:
\begin{equation}
\label{exp_theory}
W=W_i \cdot\mathrm{exp}\left(-{L}^{\ast }\right)= W_i \cdot\mathrm{exp}\left(-\frac{{E}_{0}^{2}}{{E}^{2}}\right)
\end{equation}
with
\begin{equation}
{E}_{0}^{2}=\frac{8\pi {\omega }_{0}^{2}{\omega }^{2}{m}^{3 / 2}{\tau}_{{i0}}}{\bar{\epsilon}\sqrt{2{\varepsilon }_{\text{g}}}}\,
\underset{{\varepsilon}^{\ast } / {\varepsilon }_{\text{g}}}{\overset{{\varepsilon }_{\text{F}} / {\varepsilon }_{\text{g}}}{\int }}\frac{{\left(2z+1\right)}^{2}}{\left({z}^{2}+z\right)^{3 / 2}}dz\,,
\label{eq:6}
\end{equation}
where $W_i$ is the probability of the single  impact ionization event. 
This equation describes well the main features of the observed photoconductivity. Indeed, as discussed above, at high electric fields the relative change of the conductivity $\Delta\sigma/\sigma\propto W$ varies exponentially with the square of the inverse radiation electric field, see Fig.~\ref{PC_Logarithmic}.
%
%
Furthermore, the above equation reveals that the characteristic electric field $E_0$ scales linearly with the radiation angular frequency $\omega$, which also corresponds to the results of the experiments, see inset in Fig.~\ref{PC_Logarithmic}. Finally we note that the observed increase of the characteristic field $E_0$ with rising temperature, see inset in Fig.~\ref{temp2}, is not surprising, showing that the generation rate of electron-hole pairs decreases with increasing temperature. Indeed, typically at higher temperatures the same relative increase of the electron temperature is obtained at higher electric fields.

\begin{figure}
	\centering
	\includegraphics[width=\linewidth]{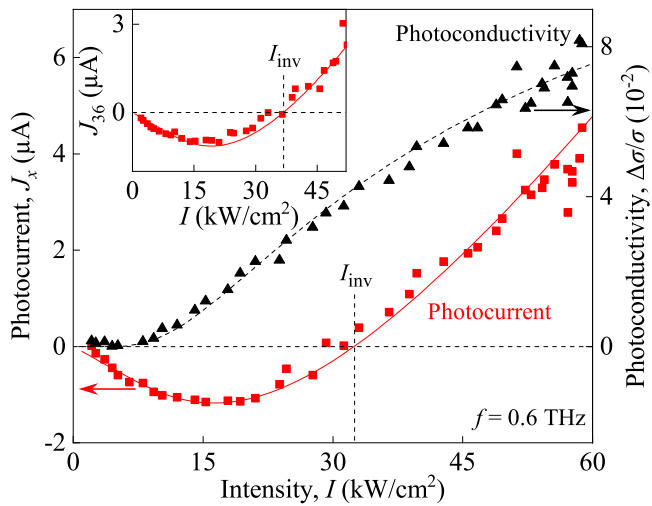}
	\caption[]{Dependences of the photoconductivity (black triangles) and the photocurrent (red squares) excited by linearly polarized radiation with a frequency of $f=0.6$\,THz on the intensity. Signals were measured between contacts 2 and 5. Dashed black line shows fit after Eq.~\eqref{exp_eqn}. 
	Here the fitting parameters $A=0.15$ and $I_0=41$\,kW/cm$^2$ obtained from the photoconductivity measurements, see Figs.~\ref{PC_Intensity} and \ref{PC_Logarithmic}, are used. Solid red line shows a fit according to 
	Eq.~\eqref{PGE_total} [also see Eqs.~\eqref{PGE_cb} and \eqref{PGE_vb}]. Here, $A$ and $I_0$ are taken from the photoconductivity fit, $n=3\times 10^{11}$\,cm$^{-2}$ from transport measurements and the parameters $ \chi_c=3.4\times 10^{-13}$\,$\upmu$A$\cdot$cm$^4$/kW and $\chi_v=8.2\times 10^{-12}\,\upmu$A$\cdot$cm$^4$/kW are obtained from fitting. At low intensities the photocurrent varies linearly with the intensity, while the photoconductivity signal is negligibly small. At higher intensities the photocurrent starts to show nonlinear behavior as the photoconductivity increases superlinearly. The inset shows the dependence of the diagonal photocurrent between contacts 3 and 6 excited by linearly polarized radiation with a frequency of $f=0.6$\,THz on the intensity. The solid line shows a fit according to Eq.~\eqref{PGE_total} with the fitting parameters $ \chi_c=2.7\times 10^{-13}\,\upmu$A$\cdot$cm$^4$/kW and $\chi_v=6\times10^{-12}\,\upmu$A$\cdot$cm$^4$/kW. It is seen that the inversion intensity $I_{\text{inv}}$ increases in this measurement direction.}
	\label{PVPC_Fit}
\end{figure}

\subsection{Photocurrent}

Now we discuss the observed photocurrent and its nonlinearity at high intensities. Instead of the quasi-momentum ${\bm p}$, we use the wave vector ${\bm k} = {\bm p}/\hbar$ for the description of the charge carrier states. We start with the analysis of the photocurrent formation mechanism. As addressed above it is mainly caused by the LPGE, in the experiment under study the circular photocurrent contribution is unessential. Two mechanisms can contribute to the LPGE current generation: 

(i) the shift photocurrent which arises from the second-order interaction with the electromagnetic wave and is related to the carrier displacement in the real space under direct or indirect optical transitions \cite{Wittmann2010,Belinicher1982,Sturman1992,Sipe2000,Ivchenko2005,Golub2011,Young2012,Bajpai2018}; and 

(ii) the ballistic photocurrent, which arises beyond the Born approximation in calculating the optical matrix element.  In the latter case, the carrier transition rate is asymmetric in the ${\bm k}$ space, and the photocurrent is stabilized by the momentum scattering time $\tau$ \cite{Sturman1992,Ivchenko2005,Belinicher1980}. 

In general, both contributions appear due to the same asymmetries of elementary processes 
of photon absorption and charge carrier scattering by phonons or defects and are comparable in order of magnitude. Here we focus the attention on the shift contribution and employ a model which describes solely the LPGE.

We consider the photocurrent generated under intraband indirect (Drude-like) optical transitions in the lowest conduction band ($c$) and the highest valence band ($v$). Taking into account that in our experiment at high intensities both electrons and holes are simultaneously present in the sample, the shift photocurrent density is a sum ${\bm j}^{(c)} + {\bm j}^{(v)}$ of the electron and hole contributions. The shift photocurrent density ${\bm j}^{(l)}~~ (l=c,v)$ is given by the following microscopic equation \cite{Ivchenko2005}  
\begin{equation} \label{shift0}
{\bm j}^{(l)}  = 2 q_l \sum_{{\bm k}'{\bm k}} W_l ({\bm k}', {\bm k}) {\bm R}_l({\bm k}', {\bm k}) \:,
\end{equation}
where $ W_l ({\bm k}', {\bm k})$ is the probability rate of the optical transitions from the free-carrier state $|l {\bm k}\rangle$ to $|l {\bm k}' \rangle$,  $q_l$ is the charge of free carriers, $q_c = e$ for an electron and $q_v = -e = |e|$ for a hole,  ${\bm R}_l({\bm k}', {\bm k})$ is the elementary shift under an indirect one-photon transition, and the factor of 2 accounts for the spin degeneracy. 

The intraband photon absorption process involves participation of a static defect or  a phonon. In the limit $\omega \gg \tau^{-1}$ the transition rate is calculated  using Fermi's golden rule for indirect transitions.
For static-defect-assisted transitions this rule reads
\begin{equation} \label{43-2}
 W_l ({\bm k}', {\bm k}) = \frac{2 \pi}{\hbar} N_d
|M_{l {\bm k}',l{\bm k}}|^2 (f_{lk} - f_{lk'}) \delta (E_{lk'} - E_{lk} - \hbar \omega)\:,
\end{equation}
with the compound matrix element being
\begin{equation} \label{indirectmm}
M_{l {\bm k}',l{\bm k}} =  \frac{ V_{l}({\bm k}) - V_{l}({\bm k}') }{ \hbar \omega}\ U_l ({\bm k}', {\bm k})\:.
\end{equation}
Here $U_l ({\bm k}', {\bm k})$ is the matrix element of scattering by a defect,  $m_l$ is the effective mass $m_c$ or $m_v$, $N_d$ is density of scattering defects,  $E_{lk} = \hbar^2 k^2/(2 m_l)$, $f_{lk}$ is the electron distribution function, $V_l({\bm k})$ is the matrix element of carrier-light interaction, the difference of the matrix elements can be presented as
\begin{equation} \label{difference}
V_l({\bm k}) - V_l({\bm k}') = \frac{q_l \hbar A_0}{m_l c} {\bm e}\cdot\left( {\bm k}' - {\bm k}\right)\:,
\end{equation}
where $A_0$ is the scalar amplitude of the light vector potential.
While deriving Eq.~(\ref{indirectmm}) we took into account the energy conservation law $E_{l{\bm k}'} - E_{l{\bm k}}= \hbar \omega$.
Using the theory of the shift LPGE \cite{Golub2011} derived for the indirect optical transitions, 
we obtain for the elementary shift in Eq. (\ref{shift0})
\begin{eqnarray} \label{shift}
{\bm R}_l ({\bm k}', {\bm k}) &=& - \frac{{\rm Im} \left[ U^*_l ({\bm k}', {\bm k}) \left( {\bm \nabla}_{{\bm k}'} + {\bm \nabla}_{\bm k} \right) U_l ({\bm k}', {\bm k})\right] }
{\left\vert U_{l {\bm k}',l{\bm k}} \right\vert^2}\nonumber\\ &+& {\cal A}_{l{\bm k}'} - {\cal A}_{l{\bm k}}\:, 
\end{eqnarray}
where ${\cal A}_{l{\bm k}}$ is the Berry connection. In the steady-state regime of photoexcitation the contributions of the Berry connections from all the processes cancel each other.
Because of the hermiticity of the operator of carrier-defect interaction, one has
\begin{equation} \label{Hermicity}
U_l^* ({\bm k}', {\bm k}) = U_l ({\bm k}, {\bm k}')\:.
\end{equation}
It is worth to note that if the matrix element $U_l ({\bm k}', {\bm k})$ depends only on the difference ${\bm k}' - {\bm k}$ the shift (\ref{shift}) vanishes. Therefore we have to make allowance for the dependence of $U_l ({\bm k}', {\bm k})$ not only on the difference but also on the sum ${\bm k}' +\ {\bm k}$. For brevity, we shall leave aside the general case and discuss a special model leading to a nonvanishing shift (\ref{shift}). Namely, we expand the matrix element (\ref{Hermicity}) up to the second order, as follows,
\begin{equation}
 \label{extension}
U_l ({\bm k}', {\bm k}) = U^{(l)}_0 + {\rm i} U^{(l)}_{1,\alpha} (k'_{\alpha} - k_{\alpha})  
+\ U^{(l)}_{2, \beta\gamma}(k'_{\beta} k'_{\gamma} + k_{\beta}k_\gamma) \:,
\end{equation}
where $\alpha, \beta, \gamma = x,y$ and $U^{(l)}_{2, \beta\gamma} = U^{(l)}_{2, \gamma \beta}$. The shift contributing to the photocurrent is given by
\begin{equation}
\label{shift2}
\left| U_l ({\bm k}', {\bm k}) \right|^2 R_{l \beta}({\bm k}', {\bm k}) = 2 U^{(l)}_{1,\alpha} U^{(l)}_{2, \beta\gamma} (k'_{\alpha} - k_{\alpha})  (k^{\prime}_{\gamma} + k_{\gamma})\:.
\end{equation}

We omit intermediate calculations and give final formulas for the photocurrents. 
For the linearly polarized light (${\bm E} \parallel x$) passed through the $\lambda/4$ plate 
the result reads
\begin{eqnarray} \label{platefinaver}
j_{l,x} =   \left(\chi^{(0)}_{l,x}  + \chi^{(1)}_{l,x} \frac{1 + \cos{4 \varphi}}{2} + \chi^{(2)}_{l,x}\frac{\sin{4 \varphi}}{2} \right)
 n_l I , \nonumber\\
j_{l,y} = \left( \chi^{(0)}_{l,y}  - \chi^{(1)}_{l,y}\frac{1 + \cos{4 \varphi}}{2} + \chi^{(2)}_{l,y} \frac{\sin{4 \varphi}}{2}\right)
n_l I, 
\end{eqnarray}
in agreement with Eq.~(\ref{N1}). Here $n_l$ are the electron and holes densities, $n$ for $l=c$ and $p$ for $l=v$, coefficients $\chi^{(l)}_{i,\alpha}$  are given by
\begin{eqnarray} \label{chiI}
&&\chi^{(0)}_{l,\alpha}= 2 \frac{\eta_l(I)}{n_l}\frac{q_l m_l}{\hbar^2} \frac{U^{(l)}_{2,\alpha\gamma}U^{(l)}_{1,\gamma}}{U_0^{(l)2}}\:,\nonumber\\
&&\chi^{(1)}_{l,\alpha} = \frac{\eta_l(I)}{n_l}\frac{q_l m_l}{\hbar^2} \frac{U^{(l)}_{2,\alpha\gamma}U^{(l)}_{1,\gamma}(2\delta_{\alpha\gamma}-1)}{U_0^{(l)2}}\:,\nonumber\\
&&\chi^{(2)}_{l,\alpha}=\frac{\eta_l(I)}{n_l} \frac{q_l m_l }{\hbar^2}\frac{U^{(l)}_{2,\alpha\beta}U^{(l)}_{1,\gamma}(1-\delta_{\beta\gamma})}{U_0^{(l)2}}\:,
\end{eqnarray}
where $\eta_l(I)$  is the absorbance defined by
\begin{equation}
\eta_l(I)=W_l(I) \frac{\hbar \omega} {I}\:,
\end{equation}
and $W_l(I) $ is the absorption rate per unit area contributed by the charge carriers $l$. Because the electron and hole densities are nonequilibrium the absorbance is dependent on the light intensity.
While deriving Eqs.~(\ref{chiI}) we assumed the ${\bm k}$-dependent terms in the expansion (\ref{extension}) to be smaller as compared with the first term $U^{(l)}_0$. The analysis shows that in the experimental conditions related to Fig.~\ref{PVPC_Fit} the ratio $\eta_l(I)/n_l$ is independent of $n_l$ and the introduction of concentration as a factor in Eqs.~(\ref{platefinaver}) is justified.

The  (013)-oriented HgTe quantum-well structure has no point-group symmetry elements except for the trivial identity operation $C_1$. As a result, the structure has no special in-plane axes, the symmetry impose no restrictions on the coefficients $U_{1,\alpha}^{(l)}$ and $U_{2,\alpha}^{(l)}$ in Eq.~(\ref{extension}) and that, in Eq.~(\ref{platefinaver}), the sets of coefficients $\chi^{(i)}_{c,\alpha}$ for the conduction electrons and coefficients  $\chi^{(i)}_{v,\alpha}$ for the holes are linearly independent. This means, particularly, that the electron and hole photocurrents are not necessarily parallel to each other, see the final remarks at the end of the Section. Equations~\eqref{platefinaver} describe well the experimentally observed polarization dependences
of the photocurrent $J \propto j$ as shown in Fig.~\ref{bulkedge}(c). To fit experimental data we consider polarized radiation with electric field vector $\bm E$ polarized along the $x$ axis and the photosignal measured along the $x$-direction, between the contacts 2 and 5.

Now we discuss the observed nonlinearity and dynamical sign change of the photocurrent. The process of light impact ionization creates electron-hole pairs resulting in an increase of the photogalvanic effect in the conduction band described by
\begin{equation}
\label{PGE_cb}
j^{(c)}_{x}=\chi_c (n_0+\Delta n) I\:,
\end{equation}
as well as in the generation of holes and appearance of a hole photogalvanic current described by
\begin{equation}
\label{PGE_vb}
j^{(v)}_{x}=\chi_v \Delta p I\:.
\end{equation}
Here $n_0$ is the equilibrium electron density, $\Delta n=\Delta p \propto \exp \left( -I_0/I \right)$ are the densities of electrons and holes generated by the light impact ionization, $$\chi_c=\kappa(\chi_{c,x}^{(0)}+\chi_{c,x}^{(1)})\:, \chi_v=\kappa(\chi_{v,x}^{(0)}+\chi_{v,x}^{(1)})\:,$$ $\kappa$ is the parameter that relates the measured photocurrent to the current density (\ref{platefinaver}) and is determined by the sample geometry and the laser spot position and diameter.
With increasing the intensity the hole density increases and, in case $|\chi_v| > |\chi_c|$, the valence band contribution of the photogalvanic effect may become comparable to and even exceed that of the conduction band. 

Fitting our data by the sum 
\begin{equation}
\label{PGE_total}
j_{x}=j^{(c)}_{x}+j^{(v)}_{x}
\end{equation}
and assuming the contributions of electrons and holes to have opposite signs, we have obtained a good agreement taking the value of $\chi_v$ by an order of magnitude larger than $\chi_c$, see Fig.~\ref{PVPC_Fit}. In this figure the data are fitted using only one adjustable parameter $\chi_v/\chi_c$. The coefficient $A$ determining in Eq.~(\ref{exp_eqn}) the number of photogenerated carriers is obtained from the photoconductivity data, see the dashed line in Fig.~\ref{PVPC_Fit}. The free carrier densities are calculated from the magnetotransport measurements, and $\chi_c$  from the data at low intensities where the nonlinearity does not play a role, see squares in Fig.~\ref{PVPC_Fit}, and the hole photocurrent is absent. The larger value of the inversion intensity for the linear polarization, $I_{\rm inv}^{\text L} > I_{\rm inv}^{\text C}$, compare Figs.~\ref{bulkedge}(a) and \ref{bulkedge}(b), is attributed to the inequality 
$$\left|\frac{\chi^{(0)}_v}{\chi^{(0)}_c}\right| > \left|\frac{\chi^{(0)}_v + \chi^{(1)}_v}{\chi^{(0)}_c + \chi^{(1)}_c}\right|\:.$$

The possibility of the valence band contribution to be indeed larger than that of the conduction band is also supported by numerical computation carried out for a phonon-involved indirect optical transitions. Calculations of the shift photocurrent for free charged carriers described by using the Bernevig--Hughes--Zhang model  \cite{Bernevig2006} with indirect optical transition processes involving scattering by acoustic phonons in the (013)-oriented HgTe QWs~\cite{Olbrich2013}, show that  the hole photocurrent can exceed the electron contribution by an order of magnitude. Details of this calculations are out of focus of the current paper.

Finally, it is important to note that a difference between the inversion points for the current measured in the $x$- and diagonal directions, Fig.~\ref{PVPC_Fit} and the inset in the figure, is caused by the non-parallel directions of the electron and hole photocurrents predicted by Eqs.~(\ref{platefinaver}). In agreement with the prediction for the $C_1$ symmetry, the interplay of the differently oriented vectors ${\bm J}^{(c)}$ and ${\bm J}^{(v)}$ yields different inversion points for different pairs of contacts. A larger value of the inversion intensity $I_{\rm inv}$ for the diagonal contacts is attributed to a smaller ratio $|\chi_v/\chi_c|$ in the $y$-direction as compared with that in the $x$-direction. At last but not at least we note that the observed shift of the inversion intensity $I_{\text{inv}}$ to higher intensity due to temperature increase, see inset in Fig.~\ref{temperature}, is in agreement with the reduction of the nonlinearity confirmed by the photoconductivity data, see inset in Fig.~\ref{temp2}.

\section{Summary} 
\label{summary}
To summarize, our experiments demonstrate that excitation of doped HgTe-based QW structures exhibiting a noninverted band order results in light impact ionization, despite the fact that the Fermi energy exceeds the energy of the forbidden gap. This effect results in strongly nonlinear photogalvanic current exhibiting a dynamic sign inversion and positive photoconductivity due to the electron-pair generation. The developed theory of the observed phenomena describes well all characteristic features of the nonlinear opto-electronic phenomena. As for the future work, we expect that experiments on QW structures with Fermi energy in the forbidden gap should allow to examine photoreponse under conditions with suppressed impact ionization process and may explore the Floquet states formation. Such experiments require structures with a semitransparent gate robust to high power terahertz radiation.

\section{Acknowledgments}
\label{acknow}We thank I.A. Dmitriev and S.N. Danilov for helpful discussions.
The support from the DFG priority program SFB 1277 (project A04), the Elite Network of Bavaria (K-NW-2013-247) and the Volkswagen Stiftung Program is gratefully acknowledged. G.V. Budkin acknowledges the support of “BASIS” foundation.

\section{Appendix}
	
\subsection{Electron-optical phonon collision integral }

We start with description of the loss term of collision integral corresponding to electron transitions out of the state with momentum $\bm p$ due to spontaneous phonon emission. We denote as $\bm k=\bm p/\hbar$ and $\bm k'=\bm k-\bm q_{\parallel} $, correspondingly, the electron wave vectors before and after emission of phonon of wave vector $\bm q=\bm q_\parallel +\bm q_{z} $, where $\bm q_\parallel $ and $\bm q_{z} $ are the in-plane and $z$-components of vector $\bm q $, respectively (here the momentum conservation law is taken into account). In these terms,
\begin{equation}
	\begin{gathered}
	\label{I}
	St_{-}^{\rm ph} \{ f\} =\frac{2\pi }{\hbar } \int \left|M_{\bm q} \right| ^{2} \delta (\varepsilon_{k}-\varepsilon_{k'}-\varepsilon_{0})\\  \times \theta (\varepsilon_{k} -\varepsilon_{0} )f(\bm k,t)\left[1-f(\bm k',t)\right]\frac{d^{2} k'dq_{z} }{(2\pi )^{3} }\:,
	\end{gathered}
\end{equation}         
where $M_{\bm q}$ is the matrix element of electron-phonon interaction \cite{Anselm1981}: 
%
\begin{equation} 
\begin{split}  
\label{II}                 
M_{\bm q} =\sqrt{\frac{4\pi e^{2} \eta \omega _{0} }{\bar{\epsilon}} } \frac{J(q_{z} )}{\sqrt{q_{\left\| \right. }^{2} +q_{z}^{2} } } \\ J(q_{z} )=\int \psi^2 (z) {\rm e}^{ {\rm i} q_{z} z} dz \:.  
\end{split}
\end{equation}                     
Here the Heaviside step function $\theta (\varepsilon -\varepsilon _{0} )=1$ at $\varepsilon \ge \varepsilon _{0} $ and $\theta (\varepsilon -\varepsilon _{0} )=0$ at $\varepsilon <\varepsilon _{0} $.

 The 2D electron wave function $\psi (z)$ decays on a scale of the quantum well width $W$. Substituting Eq.~\eqref{II} in Eq.~\eqref{I} and  integrating over $q_{z} $ and over the modulus of $\bm k'$ for $q_{\parallel} W\ll1$ (only the lowest energy level is populated) yield
\begin{equation}
\label{IIIa}
\begin{gathered}
St_{-}^{\rm ph}\{ f\} =\frac{2\pi ^{2} e^{2} \varepsilon_{0} g(\varepsilon -\varepsilon _{0} )}{\bar{\epsilon }}\\ \times \int\limits _{0}^{2\pi }\frac{f(\varepsilon ,\varphi ,t)\left[1-f(\varepsilon -\varepsilon _{0} ,\varphi',t)\right]\theta (\varepsilon -\varepsilon _{0} )}{\sqrt{p^{2} (\varepsilon )+p^{2} (\varepsilon -\varepsilon _{0} )-2p(\varepsilon )p(\varepsilon -\varepsilon_{0} )\cos (\varphi -\varphi')} } \frac{d\varphi'}{2\pi } .
\end{gathered}
\end{equation}
Similarly we obtain: 
\begin{equation}
\label{IIIb}
\begin{gathered}
St_{+}^{\rm ph} \{ f\} =\frac{2\pi ^{2} e^{2} \varepsilon _{0} g(\varepsilon +\varepsilon _{0} )}{\bar{\epsilon }} \\ \times \int\limits _{0}^{2\pi }\frac{\left[1-f(\varepsilon ,\varphi ,t)\right]f(\varepsilon +\varepsilon _{0} ,\varphi ',t)}{\sqrt{p^{2} (\varepsilon )+p^{2} (\varepsilon +\varepsilon _{0} )-2p(\varepsilon )p(\varepsilon +\varepsilon _{0} )\cos (\varphi -\varphi ')} } \frac{d\varphi '}{2\pi }.
\end{gathered}
\end{equation}

\subsection{Derivation of balance equations}

We look for a solution of the kinetic equation \eqref{eq:1} in the form of series
\begin{equation}
\label{IV}
f(\varepsilon ,\varphi ,t)=\sum\limits_{n=0}^{\infty }f_{n}  (\varepsilon ,t)\cos (n\varphi ).
\end{equation}
Substituting this series in \eqref{eq:1}, we obtain an infinite chain of equations for the functions $f_{n} (\varepsilon ,t)$. For high rate of collisions with impurities, the distribution function is almost isotropic, which allows us to consider only the first two terms of the series \eqref{IV}. The corresponding equations for the functions $f_{0} (\varepsilon ,t)$ and $f_{1} (\varepsilon ,t)\ll f_{0} (\varepsilon ,t)$ have the form:

\[
\begin{gathered}
\frac{\partial f_{0} }{\partial t} +\frac{eE v(\varepsilon  )\cos \omega t}{2} \frac{\partial f_{1} }{\partial \varepsilon } +\frac{eE\cos \omega t}{2p(\varepsilon )} f_{1} = \\ = -St_{-}^{\rm ph} \left\{ f_{0} \right\} +St_{+}^{\rm ph} \left\{ f_{0} \right\} ,                  
\end{gathered}
\]

\begin{equation}      
\label{V} 
\frac{\partial f_{1} }{\partial t} +eEv(\varepsilon)\cos \omega t\frac{\partial f_{0} }{\partial \varepsilon } =-\frac{1}{\tau _{i}(\varepsilon) } f_{1}.
\end{equation}                                            
In accordance with the assumptions made, in the second of these equations we neglect the contribution of phonon scattering to relaxation of the first harmonic, i.e. to the transport relaxation time. At frequencies exceeding all inverse scattering times, the contribution of the terms oscillating with time to the function $f_{0} (\varepsilon ,t)$ is small, i.e. $f_{0} (\varepsilon ,t)\approx f_{0} (\varepsilon )$. Therefore, from the second equation it follows that $f_{1} (\varepsilon ,t)=f_{c} (\varepsilon )\cos \omega t+f_{s} (\varepsilon )\sin \omega t$. Substituting these functions in Eq.~\eqref{V}, after simple transformations we obtain equations for $f_{0} (\varepsilon )$ in the form
\begin{equation}
\label{VIa}
\begin{gathered}
-\frac{1}{g(\varepsilon )} \frac{\partial }{\partial \varepsilon } \left[g(\varepsilon )D(\varepsilon )\frac{\partial f_{0} }{\partial \varepsilon } \right]=\frac{1}{g(\varepsilon )} \left[F(\varepsilon +\varepsilon _{0} )-F(\varepsilon )\right], \,\, \varepsilon \ge \varepsilon _{0} ,      
\end{gathered}    
\end{equation}                              

\begin{equation}
\begin{gathered}
\label{VIb}
-\frac{1}{g(\varepsilon )} \frac{\partial }{\partial \varepsilon }\left[g(\varepsilon )D(\varepsilon )\frac{\partial f_{0}}{\partial\varepsilon}\right]=\frac{1}{g(\varepsilon )} F(\varepsilon +\varepsilon _{0} ), \,\, \varepsilon<\varepsilon _{0} .
\end{gathered}
\end{equation}      
Here
\begin{equation}
	\label{VII}
	\begin{gathered}
	F(\varepsilon )=\frac{4\pi e^{2} \varepsilon _{0} g(\varepsilon -\varepsilon _{0} \, )g(\varepsilon )}{p(\varepsilon )}\ \mathbf{K} \left(\frac{p{\kern 1pt} (\varepsilon -\varepsilon _{0} )\, }{p(\varepsilon )} \right)\\ \times f_{0} (\varepsilon )[1-f_{0} (\varepsilon -\varepsilon _{0} )],
	\end{gathered}
\end{equation} 
$\mathbf{K} (x)$ is the complete elliptic integral of the first kind, which equals $\pi/2$ at $x=0$ and slowly increases with increasing $x$. Below we will replace it with $\pi/2$. Assuming now that the distribution function changes only slightly on a scale of the order of the phonon energy, we expand the right-hand side of Eq.~\eqref{VIa} in $\varepsilon _{0}$, and keep the first non-vanishing term only.  This gives 
\[
-\frac{1}{g(\varepsilon )} \frac{d}{d\varepsilon } \left[g(\varepsilon )\cal{J}(\varepsilon )\right]=0,\]

\[ {\cal{J}}(\varepsilon )=D(\varepsilon )\frac{\partial f_{0} }{\partial \varepsilon } +\frac{\varepsilon _{0} }{\tau _{\rm ph} (\varepsilon )} f_{0} (\varepsilon )\left[1-f_{0} (\varepsilon )\right],
\] 
\begin{equation}
\label{VIII}
D(\varepsilon )=\frac{e^{2} E^{2} v^ 2  (\varepsilon )}{4\omega ^{2} \tau _{i} (\varepsilon )}   ,\quad \frac{1}{\tau _{\rm ph} (\varepsilon )} =\frac{2\pi ^{2}\varepsilon _{0} e^{2} g(\varepsilon )}{\bar{\epsilon } p(\varepsilon )} .
\end{equation}

The quantity $\cal{J}(\varepsilon )$ has the meaning of particle flux in the energy space, with the first term describing the diffusive heating of the electron gas, and the second term - the relaxation of energy due to  phonon emission. From the condition of vanishing total flux, one obtains Eq.~\eqref{exp_eqn}
of the main text. Further, Eq.~\eqref{eq:2}
of the main text is obtained from Eq.~\eqref{VIb}, where one needs to put
\begin{equation}
\label{IX}
F(\varepsilon )=\frac{2\pi ^{2} e^{2} \varepsilon _{0} g(\varepsilon -\varepsilon _{0} \, )g(\varepsilon )}{p(\varepsilon )} f_{0} (\varepsilon )[1-f_{0} (\varepsilon -\varepsilon _{0} )] ,
\end{equation}
so that
\begin{equation}
\label{X}
\frac{1}{\tau _{\rm ph}^{+} (\varepsilon )} =\frac{2\pi ^{2} e^{2} \varepsilon _{0} g(\varepsilon -\varepsilon _{0})g(\varepsilon )}{\bar{\epsilon }p(\varepsilon )} .
\end{equation}

\FloatBarrier
\bibliography{ref}

\begin{thebibliography}{47}
\bibitem[{foo()}]{footnotescattering}
\bibinfo{note}{Note that the matrix element of electron scattering by short
  range impurities does not depend on energy. Thus energy dependence of the
  scattering rate, being proportional to the density of states (DOS) is solely
  determined by the energy dependence of the DOS}.
\end{thebibliography}

\end{document}